\documentclass[aps,prl,twocolumn,groupedaddress,showpacs]{revtex4}

\usepackage{graphics}
\usepackage{graphicx}
\usepackage{dcolumn}
\usepackage{bm}
\usepackage{amsmath,amssymb}

\newcommand{\de}{\partial}

\newcommand{\eq}[2]{\begin{equation} \label{#1} #2 \end{equation}}

\newcommand{\pv}{\mathbb{P}\int_{-\infty}^{+\infty}}

\newcommand{\etal}{{\em et al.}}

\newcommand{\schr}{Schr\"odinger }

\begin{document}

\title{Negative frequencies get real: a missing puzzle piece in nonlinear optics}

\author{Matteo Conforti$^{1}$, Andrea Marini$^{2}$, Daniele Faccio$^{3}$ and Fabio Biancalana$^{2,3}$}
\email{f.biancalana@hw.ac.uk}
\affiliation{$^1$CNISM, Dipartimento di Ingegneria dell'Informazione, Universita' di Brescia, 25123 Brescia, Italy\\
 $^2$Max Planck Institute for the Science of Light, G\"unther-Scharowsky str. 1, 91058 Erlangen, Germany\\
 $^3$School of Engineering and Physical Sciences, Heriot-Watt University, EH14 4AS Edinburgh, UK}

\date{\today}

\begin{abstract}
Motivated by recent experimental results, we demonstrate that the ubiquitous pulse propagation equation based on a single generalized nonlinear \schr equation is incomplete and inadequate to explain the formation of the so called negative-frequency resonant radiation emitted by optical solitons. The origin of this deficiency is due to the absence of a peculiar nonlinear coupling between the positive and negative frequency components of the ultrashort pulse spectrum during propagation, a feature that the slowly-varying envelope approximation is unable to capture. We therefore introduce a conceptually new model, based on the envelope of the analytic signal, that takes into account the full spectral dynamics of all frequency components, is prone to analytical treatment and retains the simulation efficiency of the nonlinear \schr equation. We use our new equation to derive from first principles the phase-matching condition of the negative-frequency resonant radiation observed in previous experiments.
\end{abstract}


\maketitle


\paragraph{Introduction ---} 
The study of supercontinuum generation (SCG), i.e. the explosive broadening of the spectrum of an intense and short input pulse due to nonlinear effects in a medium, typically an optical fiber or a bulk crystal, is an active area of research since its first discovery in 1970 \cite{alfano}, due to its many applications in metrology and device characterization \cite{dudley,agrawalbook}.

Constructing a theory of SCG has proved to be crucial in order to understand and control the dynamics of pulses in optical fibers \cite{russell}. Such theory is based on the so-called generalized nonlinear \schr equation (GNLSE), an enhanced version of the integrable nonlinear \schr equation \cite{dudley}. The GNLSE, based on the concept of slowly varying envelope approximation (SVEA) of the electric field, is paradigmatic in nonlinear optics, and has been extremely successful in explaining most of the features of SCG \cite{dudley,agrawalbook}. One of the most successful predictions was the emission of dispersive waves from optical solitons, which are phase-matched at specific wavelengths, usually referred to as resonant radiation (RR) or Cherenkov radiation \cite{akhmediev,fission,biancalana,skryabinscience}. RR contributes significantly to the formation of SCG spectra and can have many applications especially when using photonic crystal fibers \cite{russell,saleh1}.

However, recent experiments have revealed that new resonant frequencies [referred to as negative -frequency resonant radiation (NRR)] can be emitted by solitons, which are not predicted by any GNLSE formulation \cite{rubino1,confortitrillo}. Such frequencies can be numerically predicted by using the so-called unidirectional pulse propagation equation (UPPE, \cite{uppe}), which includes only forward propagating waves but uses the full oscillating electric field, while the phase-mathing condition has been derived heuristically \cite{rubino1,confortitrillo}. NRR has been attributed in these works to the presence of negative frequency components in the UPPE, which are absent in the GNLSE due to SVEA. However, this claim sparked some controversy in the community \cite{biancalanaphysics}, due to a lack of a solid theoretical support that could confirm or disprove the given interpretation, since this radiation could be confused with that generated by backward waves or by the conventional four-wave mixing between the soliton and copropagating radiation as in Ref. \cite{skryabinyulin}. It is also interesting to notice that, despite the fact that negative frequencies are routinely used in quantum optics \cite{quantumoptics}, quantum field theory \cite{mandlshaw} and water waves \cite{water}, in nonlinear optics there is still some resistance in accepting this concept.

In this paper we introduce a new envelope equation for a properly defined pulse envelope that is able to capture the surprising and peculiar interaction between positive and negative frequency components during the propagation of an ultrashort pulse. Such an interaction is able to generate phase-matched dispersive waves that would not exist in any model based on the conventional envelope defined when deriving the NLSE, currently referred to as NRR in the literature. We demonstrate that our new equation are easy to solve, fast to simulate and give an analytical insight into the very nature of ultrashort pulse propagation in any dielectric medium. Moreover, in this paper we also show that there are some serious deficiencies in the universally adopted equation based on the GNLSE, since the latter neglects the contribution of the cross-phase modulation between the positive and negative frequency parts of the spectrum, which gives rise to new and unexpected nonlinear phenomena that have been previously completely overlooked.

\paragraph{Definitions --- } In this section we make some important definitions that we shall use throughout the paper, following Refs. \cite{amir1,matteochi2}. The real electric field propagating in the fiber is denoted by $E(z,t)$, where $z$ is the propagation direction and $t$ is the time variable. The Fourier transform of the electric field is denoted by $E_{\omega}(z)\equiv\mathcal{F}[E(z,t)]=\int_{-\infty}^{+\infty}E(z,t)e^{i\omega t}dt$. The {\em analytic signal} of the electric field, i.e. the positive frequency part of the field, which is a complex function, is defined as $\mathcal{E}(z,t)\equiv \pi^{-1}\int_{0}^{\infty}E_{\omega}(z)e^{-i\omega t}d\omega$. The analytic signal can also be defined alternatively by using the Hilbert transform: $\mathcal{E}(z,t)=E(z,t)-i\mathcal{H}[E(z,t)]$, where $\mathcal{H}[E(z,t)]\equiv\pi^{-1}\pv dt'E(z,t')/(t-t')$, and the simbol $\pv$ indicates that the integral must be taken in the sense of the Cauchy principal value. With these definitions, the Fourier transform of the electric field can be written as the sum $E_{\omega}=[\mathcal{E}_{\omega}+(\mathcal{E}_{-\omega})^{*}]/2$ since only the positive (or negative) frequency part of the spectrum carries information, while for the same reason the electric field itself is real and is given by $E(z,t)=[\mathcal{E}(z,t)+\mathcal{E}^{*}(z,t)]/2$. The analytic signal satisfies the following requirements: $\mathcal{E}_{\omega>0}=2E_{\omega}\neq 0$, $\mathcal{E}_{\omega<0}=0$ and $\mathcal{E}_{\omega=0}=E_{\omega=0}$. Note that $(\mathcal{E}^{*})_{\omega}$ and $(\mathcal{E}_{\omega})^{*}$ are different in general and must be distinguished.

\paragraph{Derivation of envelope equation ---}  The starting point of our discussion is the so-called unidirectional pulse propagation equation (UPPE) \cite{uppe}, which is a reduction of Maxwell's equations that accounts only for the forward propagating part of the electric field:
\begin{equation}\label{uppe}
i\frac{\partial E_{\omega}}{\partial z}+\beta(\omega)
E_{\omega}+\frac{\omega}{2cn(\omega)}
P_{\rm{NL},\omega}=0,
\end{equation}
where $\beta(\omega)$ is the full propagation constant of the medium, $c$ is the speed of light in vacuum, $n(\omega)$ is the linear refractive index, and $P_{\rm{NL},\omega}(z)\equiv\chi^{(3)}\mathcal{F}[E(z,t)^3](\omega)$ is the nonlinear Kerr polarization. Particular care must be devoted to the definition of the complex envelope, since we do not want to put any limitation to the frequency extent of the signals. This aspect is overlooked in the literature, and it is taken for granted that the frequency bandwidth of the envelope is narrow with respect to the carrier wave.

The key element that we introduce here is that only a proper definition of the envelope is able to capture the correct coupling between the positive and the negative frequency parts of the spectrum. The 'envelope' we introduce here is based on the analytic signal and is defined as:
\eq{envelope2}{A(z,t)\equiv\mathcal{E}(z,t)e^{-i\beta_{0}z+i\omega_{0}t},} i.e. the frequency components of the analytic signal are 'shifted' by an amount $-\omega_{0}$. By doing this, we shift the carrier frequency of the analytic signal to zero, so that we deal with frequency detunings $\Delta\omega$ from $\omega_{0}$, and not with absolute frequencies, in analogy with the conventional definition of envelope done in many textbooks \cite{agrawalbook}. However, {\em there is a key difference} between the conventional definition of envelope (see e.g. Ref. \cite{agrawalbook}) and Eq. (\ref{envelope2}): the former is adequate only if the spectral extension of the pulse evolution is much smaller than the pulse central frequency, $\Delta\omega\ll\omega_{0}$, i.e. only under SVEA conditions, while the envelope of the analytic signal $A(z,t)$ considered here does not suffer from this limitation, and so $\textrm{ supp} \{A_{\omega}(z)\}=(-\omega_0,+\infty)$. By clearly dividing the envelope associated to the positive frequency components from that associated to the negative frequency components, we will be able to write the envelope equation that correctly describes the dynamics of pulses {\em of arbitrary duration and spectral extension}, taking into account the peculiar and non-trivial interaction between positive and negative frequencies that arises due to the nonlinear polarization.

With the above definitions, the nonlinear polarization is now written as:
\begin{widetext}
\begin{eqnarray}\label{Pnl}
 \nonumber P_{\rm NL}(z,t)=\frac{\chi^{(3)}}{8}\bigg[A^3e^{-3i\omega_0t+3i\beta_0z}+A^{*3}e^{3i\omega_0t-3i\beta_0z}
+3|A|^2Ae^{-i\omega_0t+i\beta_0z}+3|A|^2A^*e^{i\omega_0t-i\beta_0z}\bigg].
\end{eqnarray}
\end{widetext}
Due to our definition of $A$, the first (second) term in the square brackets contains only positive (negative) frequencies, and they are responsible for third harmonic generation (THG). The third and fourth terms {\em contain both positive and negative frequencies}, because the Fourier transform of $|A|^2$ has a frequency support that extends from $-\infty$ to $+\infty$.
In fact $\mathcal{F}[|A|^2](\omega)$ is the convolution between $A_{\omega}$, whose support is $(-\omega_0,+\infty)$], and $A^{*}_{\omega}$, whose support is $(-\infty,\omega_0)$. By applying the Titchmarsh convolution theorem (i.e. the support of the convolution is contained in the sum of the supports of its individual terms \cite{titchmarsh}), it immediately follows that $\textrm{ supp} \{\mathcal{F}[|A|^2](\omega)\}\subseteq(-\infty,+\infty)$. This means that, although in absence of nonlinearities positive and negative frequencies live a completely separate existence, {\em in presence of nonlinear terms they can interact and give rise to new non-linear phenomena}, especially in presence of resonant processes.
If we denote with $\mathcal{P}_{\rm NL}(z,t)$ the analytic signal for the nonlinear polarization, then its envelope $A_{p}(z,t)=\mathcal{P}_{\rm NL}e^{-i\beta_{0}z+i\omega_{0}t}$ can be expressed as:
\begin{widetext}
\begin{eqnarray} \label{Ap_ex}
A_p(z,t)=\frac{3\chi^{(3)}}{4}\left[|A|^2A+|A|^2A^*e^{2i\omega_0t-2i\beta_0z}+\frac{1}{3}A^{3}e^{-2i\omega_0t+2i\beta_0z}\right]_{+}
\end{eqnarray}
\end{widetext}
The subscript '+' indicates that only positive frequencies must be taken, i.e. $\Delta\omega>-\omega_{0}$, is a shorthand notation to indicate the positive frequency spectral filtering involved in the analytic signal, and operated in the time domain by the Hilbert transform, which is crucial in our formulation.

Finally, with all the above ingredients, one can write an equation for the analytic signal envelope $A$ which contains only positive frequencies (neglecting THG):
\begin{widetext}
\begin{eqnarray}
i\de_{\xi}A+\hat{D}(i\de_{\tau})A+\gamma\hat{S}(i\de_{\tau})\left[|A|^{2}A +|A|^{2}A^{*}e^{2i\Delta k\xi+2i\omega_{0}\tau}\right]_{+}=0, \label{gov1}
\end{eqnarray}
\end{widetext}
where $\Delta k\equiv (\beta_{1}\omega_{0}-\beta_{0})$, $\xi\equiv z$ and $\tau\equiv t-\beta_{1}z$ are the new space-time variables in the comoving frame, the dispersive operator $\hat{D}(i\de_{\tau})\equiv\sum_{m=2}^{\infty}\beta_{m}(i\de_{\tau})^{m}/m!$, $\gamma$ is the nonlinear coefficient of the medium, and $\hat{S}(i\partial_{\tau} )$ is the operator accounting for the dispersion of the nonlinearity, which is necessary to include since the equations are broadband and SVEA is not used. For our purposes it will be sufficient to perform the traditionally adopted approximation $\hat{S}(i\de_{\tau})\simeq1+i\de_{\tau}/\omega_{0}$.
Note that the field $A$ feels a dispersion given by $D(\Delta\omega)=\sum_{m=2}^{\infty}\beta_{m}\Delta\omega^{m}$ (where $\Delta\omega$ is the detuning from $\omega_{0}$) and a positive nonlinearity, while the field $A^{*}$ feels a different, 'conjugate' dispersion $-D(-\Delta\omega)\neq D(\Delta\omega)$ and a negative nonlinearity, and both fields are forward-propagating.

Equation (\ref{gov1}) is the central result of this paper. Since $A$ and $A^{*}$ carry the same amount of information, it is sufficient to consider a single equation only, because the dynamics around the positive carrier frequency ($\omega_{0}$) must be the mirror image of the dynamics around the negative carrier frequency ($-\omega_{0}$), due to the requirement that the electric field $E$ be real. The two modes $A$ and $A^{*}$ do not see each other in absence of nonlinearity, but they mutually exchange energy when the nonlinear terms are included, generating new frequencies. Since the interaction modifies the phase, new resonant nonlinear effects occur. It is possible to prove (although we omit here the nontrivial derivation) that in presence of the THG term [i.e. the third term in the right-hand side of Eq.  (\ref{Ap_ex})] the photon number is perfectly conserved, i.e. $\de_{z}\int_{-\infty}^{+\infty}|A(z,t)|^{2}dt=0$, due to the detailed balance of the energy flow from $A$ to $A^{*}$ and back. It is interesting to note that the presence of the shock operator is {\em essential} for the energy conservation, which establishes a deep and previously unknown connection between shock operator, THG and negative frequencies. In absence of THG terms, Eq. (\ref{gov1}) exhibits a very small non-conservation of photon number proportional to the missing THG energy.

\paragraph{Phase-matching between soliton and radiation --- } In order to derive phase-matching conditions between a soliton and its resonant radiations, we follow a standard procedure described in \cite{skryabinyulin}. We first pose $A(\xi,\tau)=F(\tau)e^{iq\xi}+g(\xi,\tau)$, where $F(\tau)$ is the envelope of the optical soliton, $q$ is the nonlinear mismatch and $g$ is a small amplitude dispersive wave. After substitution into Eq. (\ref{gov1}), and by taking only the fundamental and first order terms, one obtains (neglecting the shock term for simplicity):
\begin{eqnarray}
(i\de_{\xi}+\hat{D})g+\gamma F^{2}g^{*}e^{2iq\xi}+2\gamma F^{2}g&=& \nonumber \\
(\hat{D}+\frac{1}{2}\beta_{2}\de_{\tau}^{2})Fe^{iq\xi}
-\gamma F^{3}e^{2i\omega_{0}\tau+2i\Delta k \xi-iq\xi}. \label{pmequation}
\end{eqnarray} The phase-matching conditions derived from Eq. (\ref{pmequation}) are then easily found: \begin{eqnarray}
D(\Delta\omega)=q, \label{pm1} \\
D(\Delta\omega)=2\Delta k-q. \label{pm2}
\end{eqnarray}
Solving equations (\ref{pm1}-\ref{pm2}) for $\Delta\omega$ will provide all the phase-matched frequencies. In particular, Eq. (\ref{pm1}) is very well known (\cite{akhmediev,biancalana}) and corresponds to the positive-frequency RR, while Eq. (\ref{pm2}), found experimentally in Refs. (\cite{rubino1}) and heuristically in Ref. (\cite{rubino1,confortitrillo}), corresponds to the negative-frequency RR. The latter is impossible to find by using a single GNLSE based on SVEA and thus corresponds to a new feature of our envelope model Eq. (\ref{gov1}).

\begin{figure}
\includegraphics[width=8cm]{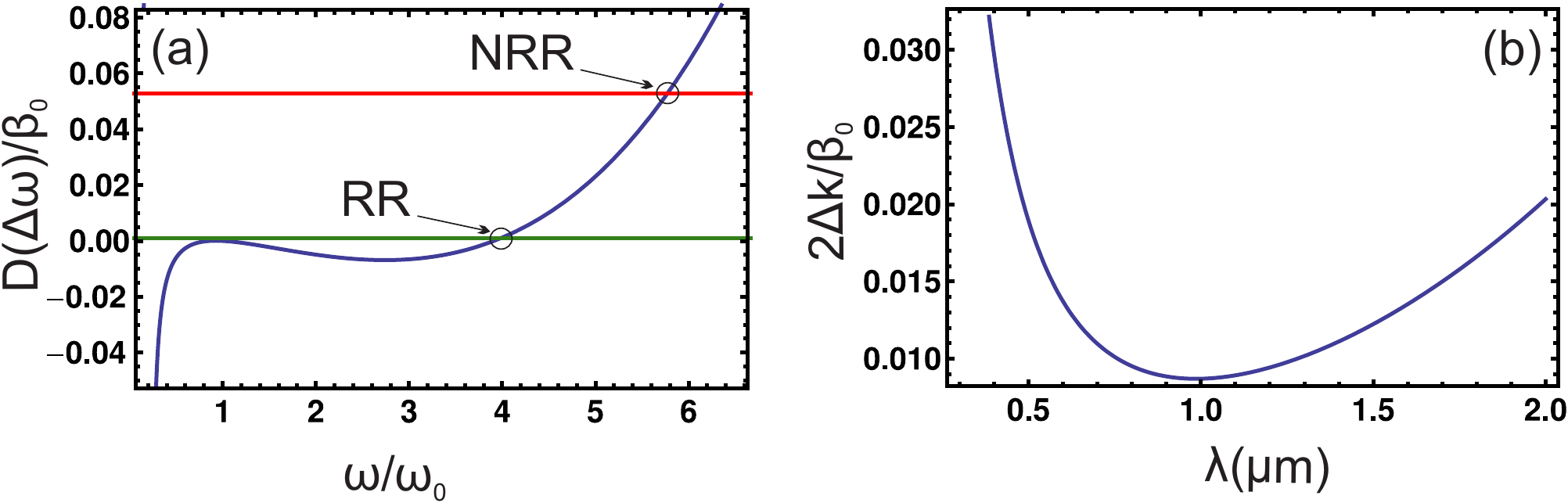}
\caption{(Color online) (a) Phase matching curve (normalized to $\beta_{0}$) derived by using Eqs. (\ref{pm1}-\ref{pm2}) in bulk silica. $q/\beta_{0}$ and $2\Delta k/\beta_{0}$ are indicated by the bottom green and the top red horizontal lines, respectively. RR and NRR are indicated by circles. (b) $\Delta k$ normalized to $\beta_{0}$ for bulk silica vs. pump wavelength. \label{fig1}}
\end{figure}

Figure \ref{fig1}(a) shows the phase-matching curve $D(\Delta\omega)$ versus pump frequency (normalized to $\beta_{0}$ and $\omega_{0}$ respectively), together with its intersection with $q$ and $2\Delta k-q$, which give respectively the RR and NRR frequencies. Figure \ref{fig1}(b) shows the normalized $2\Delta k$ versus pump wavelength, showing that in bulk silica there is an optimal pump wavelength (in the normal dispersion regime) for which the NRR would be closer to the pump frequency, an thus would have an unusually large amplitude.

\paragraph{Numerical simulations --- } In this section we support the above theory with accurate numerical simulations performed by integrating Eq. (\ref{gov1}).
\begin{figure}
\includegraphics[width=8cm]{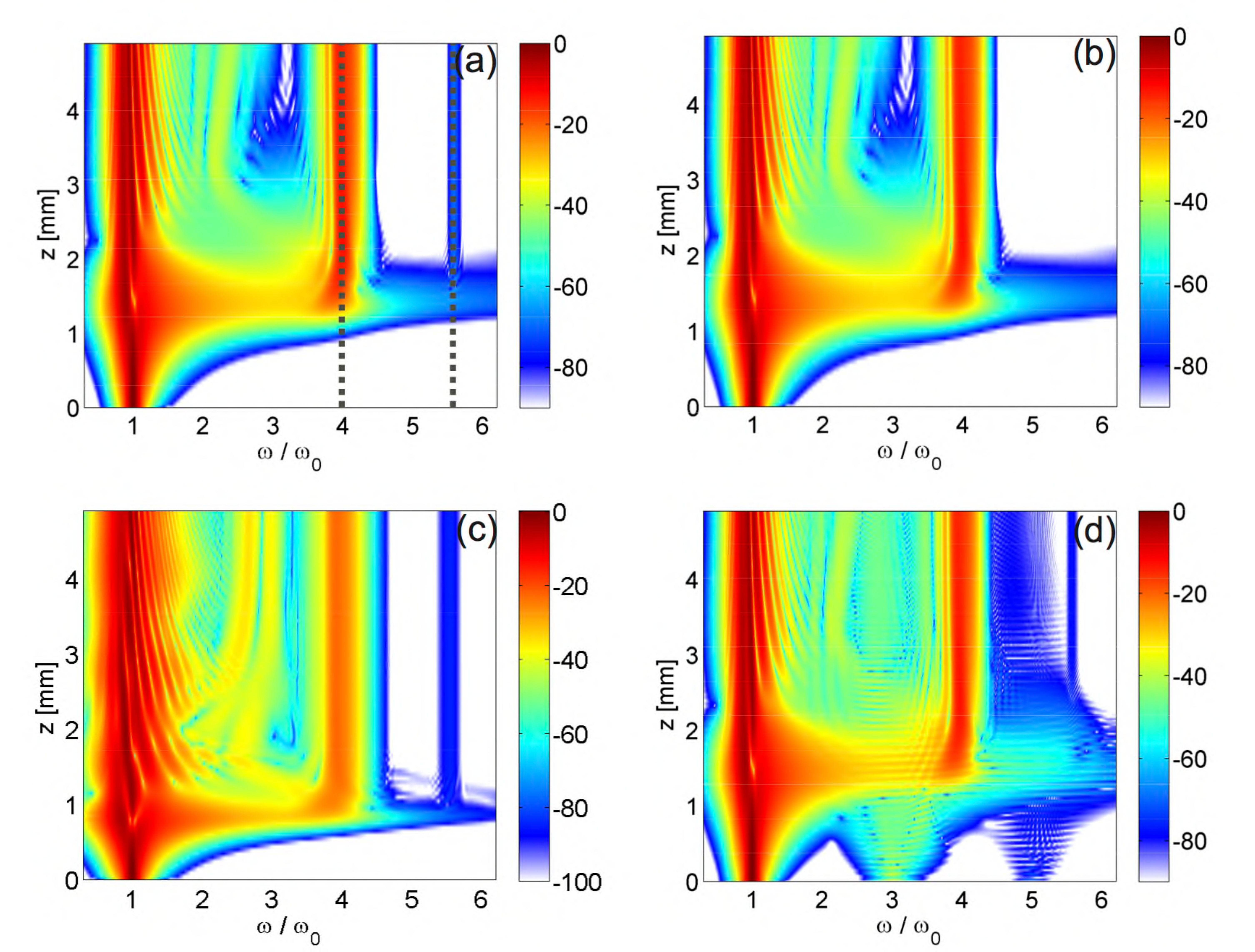}
\caption{(Color online) (a) Contour plot of the spectral evolution of a short sech pulse in bulk silica, obtained by direct simulation of Eq. (\ref{gov1}). The pulse is pumped at $\lambda_{0}=2$ $\mu$m, with a peak intensity of $1.4$ TW/cm$^{2}$ and a duration $t_{0}=15$ fsec. The formation of RR and NRR is clearly visible. Vertical black dashed lines indicate the position of the radiations as predicted by Eqs. (\ref{pm1}-\ref{pm2}), compare with Fig. \ref{fig1}(a). (b) Same as (a) when switching off the second nonlinear term in the right-hand side of Eq. (\ref{gov1}). The NRR line has completely disappeared. (c) Same as (a) when switching off the shock operator. (d) Results obtained with the UPPE of Eq. (\ref{uppe}), using the same parameters as in (a). All plots are in logarithmic scale.
\label{fig2}}
\end{figure}
In Fig. \ref{fig2}(a) we show the spectral evolution of a $15$ fs sech pulse, with peak intensity $1.4$ TW/cm$^2$ propagating in bulk silica, for a pump wavelength $\lambda_{0}=2$ $\mu$m, obtained by solving Eq. (\ref{gov1}). Both RR and NRR emissions are visible. Vertical black dashed lines indicate the predictions given by Eqs. (\ref{pm1}-\ref{pm2}), see also Fig. \ref{fig1}(a). Figure \ref{fig2}(b) shows the same as Fig. \ref{fig2}(a), when omitting the second nonlinear term in the right-hand side of Eq. (\ref{gov1}). No NRR radiation is generated in this case, showing that such radiation is indeed coming from the interaction between the positive and the negative frequency spectral components. Figure \ref{fig2}(c) shows the same simulation as in Fig. \ref{fig2}(a) but when switching off the shock term, i.e. $\hat{S}(i\de_{\tau})=1$. One can see that both RR and NRR are visible, conclusively proving that NRR is not due to the shock effect (even though the shock helps to further broaden the spectrum and thus to feed the soliton tail that excites the NRR, making it more evident). Finally, Fig. \ref{fig2}(d) shows the evolution of the pulse by solving the full-field UPPE, Eq. (\ref{uppe}), which also shows evidence of small THG. The similarity between Fig. \ref{fig2}(a) and (d) shows that our envelope model based on the analytic signal is indeed correct.

\paragraph{Discussion and conclusions ---} 
In conclusion, we have derived a novel envelope equation that correctly describes the nonlinear interaction between the positive and the negative frequency parts of the spectrum. The key concept is that the envelope function is now correctly defined in terms of the analytic signal of the electric field, therefore clearly dividing the dynamics of the negative and positive frequency parts of the spectrum, and avoiding SVEA altogether, while still retaining an envelope formulation. The interaction between positive and negative frequencies is due to the presence of cross-phase-modulation-like terms in the nonlinear polarization, the role of which we have elucidated here for the first time. By using the new equation we have been able to derive analytically the phase-matching conditions between a soliton and the positive- and negative-frequency resonant radiation emitted by it. Our theory opens up a new realm in nonlinear optics and in other areas that are described by NLSE-like equations (for instance BEC, plasmas, water waves, etc.), since it proves that conventional treatments based on GNLSE are deficient, due to the lack of the negative frequency terms. These interactions are of course present in the UPPE, which is however less transparent and less suitable for analytical treatment than Eq. (\ref{gov1}). Exciting future perspectives are represented by the inclusion of the Raman nonlinearity, which could provide additional unexplored non-linear effects that are not captured by conventional GNLSE based on SVEA.

FB is funded by the MPG (Germany), DF from the EPSRC (UK, Grant EP/J00443X/1) and from the European Research Council (FP/2007-2013) / ERC Grant Agreement n. 306559, MC from MIUR (PRIN 2009P3K72Z). We thank J. C. Travers, E. Rubino and S. Trillo for useful discussions.

\end{document}